\RequirePackage{fix-cm,amsmath}
\documentclass{svjour3}                     
\smartqed  
\usepackage{color,graphicx,amsmath,amsfonts}
\newcommand{\COMMENTED}[1]{}

\begin{document}

\title{Response functions for the two-dimensional ultracold
Fermi gas: dynamical BCS theory and beyond}

\titlerunning{Response functions for two-dimensional ultracold
Fermi gases}

\author{Ettore Vitali    \and
        Hao Shi         \and
        Mingpu Qin     \and
        Shiwei Zhang  }

\authorrunning{Vitali et al.}

\institute{Ettore Vitali and Hao Shi and Mingpu Qin and Shiwei Zhang \at  
           Department of Physics, The College of William and Mary, Williamsburg, Virginia 23187, USA\\
           \email{ettore.vitali2@gmail.com}  
}

\date{Received: date / Accepted: date}

\maketitle

\begin{abstract}
Response functions are central objects in  physics. They provide crucial information about the behavior of physical systems, and they can be directly compared
with scattering experiments involving particles like neutrons, or 
photons. Calculations of such functions starting from the many-body Hamiltonian of a physical system
are challenging, and extremely valuable.
In this paper we focus on the two-dimensional (2D) ultracold Fermi atomic gas
which has been realized  experimentally.
We present an application of
the dynamical BCS theory to obtain response functions for 
different regimes of interaction 
strengths in the 2D gas with  zero-range attractive interaction.
We also discuss auxiliary-field quantum Monte Carlo (AFQMC) methods for the calculation of
imaginary-time correlations in these dilute Fermi gas systems. Illustrative results are given and
comparisons are made between AFQMC and  dynamical BCS theory results to assess the accuracy of  the latter.

\keywords{cold atoms, response functions, dynamical BCS theory, time-dependent mean-field calculations, superfluidity,  strongly correlated fermions, auxiliary-field Quantum Monte Carlo, AFQMC}

\end{abstract}

\section{Introduction}
\label{intro}

When a collection of atoms, for example $^6$Li, is cooled to
degeneracy in an equal mixture of two hyperfine ground states,
labeled $|\uparrow\rangle$ and $|\downarrow\rangle$, 
Feshbach resonances make possible to tune the interactions by varying an external magnetic
field. 
The system exhibits an interesting phase
diagram, including a highly challenging strongly interacting regime,
the BEC-BCS crossover.

Using an highly anisotropic trapping potential, it is possible
to explore also two-dimensional (2D) geometries, which are
 intriguing, and important  in connection with, for example, 
high temperature superconductivity \cite{RevModPhys.78.17},
Dirac fermions in graphene  \cite{RevModPhys.81.109} and topological superconductors \cite{RevModPhys.83.1057}, and nuclear ``pasta'' phases \cite{PhysRevLett.90.161101} in neutron stars.

In three-dimensions (3D), there exists a unitarity point where the two-body scattering length diverges 
and the gas properties become universal in the dilute limit. The unitarity point resides in the heart of the so-called crossover regime, and provides a natural boundary for the strongly interacting BEC regime
(positive scattering length and molecular bound state) and the weakly interacting 
BCS regime (negative scattering length and
no bound state) \cite{RevModPhysStringari}.

In 2D a two-body bound state always exists, and the scattering length, $a$, is always positive. 
The relevant parameter is $\log(k_F a)$,
with $k_F$ being the Fermi momentum,
which is controlled by the particle density of the dilute gas.
%
In the a weakly interacting
BCS regime at $\log(k_F a) > 1$, the attraction between particles with
opposite ``spin'' induces a pairing similar to the one observed
in ordinary superconductors.
As the interaction strength is
increased (and correspondingly $\log(k_F a)$ decreases),  a crossover is observed leading
to a BEC regime where the Cooper pairs are so tightly bound
that the system behaves as a gas of bosonic molecules.
While both the BCS and the BEC regimes are well understood
on the grounds of the celebrated BCS theory, the crossover regime
provides a challenging example of a strongly interacting quantum
many-body system \cite{RevModPhysZwerger,RevModPhysStringari}.

An array of ground state properties in 2D have already  been  measured,
both theoretically and experimentally  \cite{PhysRevLett.112.045301,PhysRevLett.114.230401,PhysRevLett.116.045303,nature_Feld,PhysRevA.94.031606,Bauer,PhysRevA.92.023620,Klawunn20162650,PhysRevA.93.023602,Giorgini}, although much less is available in 2D compared to 3D 
systems. 
Exact calculations 
have been recently achieved using auxiliary-field quantum Monte Carlo (AFQMC) methodologies both for ground state properties  \cite{Hao-2DFG}
and excited states \cite{Ettore-2DFG}.

The main focus of this paper is the theoretical
calculation of response functions of the 2D 
Fermi gas.
The computation of these quantities is challenging, and significantly more difficult than 
the computation of static properties, since
the full spectrum of the microscopic Hamiltonian is relevant
for the response of the system to external perturbations.
The importance of computing 
such properties from
first principles can hardly be overrated: they allow for a direct
comparison with experiments, and they provide insight
into the 
behavior of the many-body physical system.
In particular the density and spin structure factors can be measured
for a cold gas in two-photon scattering experiments \cite{PhysRevLett.109.050403}, and they will be the main topic of this paper.

We discuss two theoretical approaches to compute
response functions: the dynamical BCS theory 
and AFQMC 
methods.
With dynamical BCS theory  \cite{PhysRev.139.A197,Nozieres,PhysRevA.74.042717} calculations,
we follow the approach in Ref.~\cite{PhysRevA.74.042717}, which studied the 3D Fermi gas, 
and obtain systematic results as a function of the interaction strength in the 2D gas.
We then describe our recently proposed approach 
for computing imaginary-time correlation functions with AFQMC \cite{ettoreGAP} 
and discuss how we apply it to compute the density and spin response functions.
The AFQMC approach yields 
 exact numerical results of imaginary time correlation functions for these strongly interacting systems,
which can provide 
useful benchmarks for dynamical BCS and other theoretical approaches \cite{1367-2630-18-11-113044}. We will use in particular such exact estimations
to assess the accuracy of dynamical BCS predictions.

The rest of the paper is organized as follows. In Sec.~\ref{hcgr},
we introduce the basic notations, the Hamiltonian for the system
and their regularization. In Sec.~\ref{dBCS}
 we first give a brief introduction to the
dynamic BCS theory within the framework of linear response theory and then we show the results for the dynamical structure factors. 
In Sec.~\ref{AFQMC} we introduce the auxiliary-field quantum Monte Carlo (AFQMC) methodology and discuss the details of the unbiased calculation of two-body correlation
functions in imaginary time. In Sec.~\ref{COMP}, we make a comparison of the results from dynamic BCS theory and AFQMC. We conclude
this paper in Sec.~\ref{CONC}.

\section{Hamiltonian for cold atoms and regularization}
\label{hcgr}
We model the system as a two-component ($\uparrow$
and $\downarrow$) Fermi
gas interacting through a zero-range attractive interaction 
acting only among particles with opposite ``spins'' $v_{\uparrow\downarrow}(\vec{r}_1, \vec{r}_2) = -g \delta( \vec{r}_{1} -\vec{r}_{2})$.
The basic Hamiltonian is thus:
\begin{equation}
\label{ham_r}
\hat{H} = \int d\vec{r} \sum_{\sigma}
\hat{\psi}_{\sigma}^{\dagger}(\vec{r}) \left( - \frac{\hbar^2 \nabla^2}{2m} - \mu \right) \hat{\psi}_{\sigma}^{}(\vec{r})
- g \int d\vec{r}\, \hat{\psi}_{\uparrow}^{\dagger}(\vec{r}) \hat{\psi}_{\downarrow}^{\dagger}(\vec{r}) \hat{\psi}_{\downarrow}^{}(\vec{r})
 \hat{\psi}_{\uparrow}^{}(\vec{r})\,.
\end{equation}
Introducing, as usual, a supercell $\Omega = [-L/2,L/2]^2$ with volume $V = L^2$, we can write this Hamiltonian in momentum space as:
\begin{equation}
\label{ham_mom}
\begin{split}
& \hat{H} = \sum_{\vec{k},\sigma= \uparrow,\downarrow} \left( \frac{\hbar^2 |\vec{k}|^2}{2m} - \mu \right)  \hat{c}^{\dagger}_{\vec{k}, \sigma} \, \hat{c}^{}_{\vec{k}, \sigma} \\
& - \frac{g}{V}
\sum_{\vec{k}, \vec{k}',\vec{\lambda}} \, \hat{c}^{\dagger}_{\vec{k} + \vec{\lambda}/2, \uparrow} \, \hat{c}^{\dagger}_{-\vec{k}+ \vec{\lambda}/2, \downarrow}
 \hat{c}^{}_{-\vec{k}'+ \vec{\lambda}/2, \downarrow} \, \hat{c}^{}_{\vec{k}' + \vec{\lambda}/2, \uparrow} 
 \end{split}
\end{equation}
 where, if we choose periodic boundary conditions, the momenta
 are discretized as  $\vec{k} = \frac{2 \pi}{L} \vec{n}$, $\vec{n} \in \mathbb{Z}^2$. Due to the singular nature of the interacting potential,
 which leads to divergences in summations over momentum space, some
 further regularization is needed. In this paper we will use a lattice regularization, which is particularly useful for QMC. 
We introduce a cutoff:
\begin{equation}
\sum_{\vec{k}} \longrightarrow \sum_{\vec{k} \in \mathcal{D} }
\end{equation}
where $\mathcal{D} =   [-\pi/b,\pi/b) \times [-\pi/b,\pi/b)$
is the Brillouin zone of a finite square lattice $\mathcal{L} = (b \mathbb{Z})^2 \cap \Omega$, containing
$\mathcal{N}_s = L/b \times L/b$ sites, and $b$ is the lattice parameter.
Consistently, all functions in real space are defined on
$\mathcal{L}$, and the Hamiltonian is mapped onto a lattice
Hamiltonian $\hat{H}_{\mathcal{L}} = \hat{T} + \hat{V} - \mu \hat{N}$,
which we conveniently write as follows: 
\begin{equation}
\label{ham_latt}
\hat{H}_{\mathcal{L}}  = \sum_{\vec{k} \in \mathcal{D},\sigma= \uparrow,\downarrow} \xi(\vec{k}) \,  \hat{c}^{\dagger}_{\vec{k}, \sigma} \, \hat{c}^{}_{\vec{k}, \sigma}
- g_{\mathcal{L}} \sum_{i \in \mathcal{L}} \, \hat{n}_{i,\uparrow} \hat{n}_{i,\downarrow}
\end{equation}
\COMMENTED{
 We introduce a finite square lattice containing $\mathcal{N}_s$
 sites, with lattice parameter $b$, and we 
 over $\vec{k}$

It is known that the contact interaction requires the
introduction of suitable regularization schemes to deal with infinities
arising from the Dirac delta.
In this paper we will use a lattice regularization, which is particularly
useful for QMC.
We introduce a finite square lattice containing $\mathcal{N}_s = L \times L $ sites with
lattice parameter $b$ and thus volume $V = (Lb)^2$ and we build a lattice Hamiltonian $\hat{H} = \hat{T} + \hat{V} - \mu \hat{N}$ as follows: we define a kinetic energy
term of the form:
\begin{equation}
\hat{T} -  \mu \hat{N} = \sum_{\vec{k},\sigma= \uparrow,\downarrow} \xi(\vec{k}) \,  \hat{c}^{\dagger}_{\vec{k}, \sigma} \, \hat{c}^{}_{\vec{k}, \sigma}
\end{equation}
where the momenta are defined inside the first Brillouin zone of the reciprocal lattice $\vec{k} \in [-\pi/b,\pi/b) \times [-\pi/b,\pi/b)$ and are  discretized as $\vec{k} = \frac{2 \pi}{Lb} (n_x, n_y)$, $n_{x,y}$ being integers.} 
In the interaction part the label $i$ runs over the sites of the lattice and $\hat{n}_{i,\sigma}$ denotes the spin-resolved particle density at site
$i$. 
The dispersion relation can be either a Hubbard type
$ \xi(\vec{k})  = t ( 4 - 2 \cos (k_x b) - 2 \cos (k_y b)) - \mu $, or a quadratic
dispersion $ \xi(\vec{k})  = t  ((k_x b)^2 + (k_y b)^2) - \mu$, with the
hopping constant given by $t = \hbar^2 / 2 m b^2 $.
Other forms of the dispersion can be used 
to produce  desired two-particle scattering properties, 
as long as  the value of the on-site interaction strength,
 $g_{\mathcal{L}}$, is tuned accordingly  \cite{PhysRevA.84.061602}
so that they  converge to the same continuum limit when $b \to 0$.
It can be shown that the value for 2D  is \cite{PhysRevA.86.013626}:
\begin{equation}
\label{uofeta}
\frac{g_{\mathcal{L}}}{t} = \frac{4\pi}{\ln(k_F a) - \ln(\mathcal{C} \sqrt{n})}\,,
\end{equation}
where $n=N/\mathcal{N}_s$ is particle density on the lattice, 
$k_F = \frac{\sqrt{2 \pi n}}{b}$ is the Fermi momentum,
and $\ln(k_F a) $ is the interaction strength,
containing the scattering length $a$, defined as
the position of the first node in the zero-energy $s$-wave solution
of the two-body problem . Finally, 
 $\mathcal{C}$ is a constant, whose precise
value depends on the choice of the dispersion relation (for
example $\mathcal{C} = 0.80261$ for the quadratic dispersion).

To summarize, in order to study the cold gas basic
Hamiltonian \eqref{ham_r}, we first introduce, as usual, a supercell
of finite volume $V$
and consider the Hamiltonian \eqref{ham_mom};
in order to avoid divergences, 
the use of the lattice Hamiltonian \eqref{ham_latt} is necessary
and the properties have to be extrapolated to the continuum limit $b \to 0$. Finally, extrapolation
to the thermodynamic limit $N\to \infty$ yields the results for the physical system.
We will refer to the Hamiltonian  \eqref{ham_mom}
when writing equations, but we will implicitly assume that the
lattice regularization is used and the results are extrapolated.

\section{Dynamical BCS theory}
\label{dBCS}
Our focus here is on response functions, describing the physical
response of the system to external perturbations. We will now
briefly sketch the general linear response framework, which will
allow us to introduce the dynamical BCS approximation, as well
as to emphasize the connection with dynamical structure factors,
which can be estimated from 
AFQMC calculations discussed in the next section.

\subsection{Linear response framework and dynamical BCS approximation}
Let us assume the system is in the many-body ground State $|\Psi_0\rangle$ of  \eqref{ham_mom}.
We switch on a periodic external potential $U(\vec{r},t) = \frac{1}{V} \delta U_{}(\vec{q},\omega) e^{i (\vec{q} \cdot \vec{r} - \omega t)} + c.c.$, with a well defined momentum $\vec{q}$ 
and frequency $\omega$, which couples to either the density $n$ or the spin-density  $S_z$ of the system to give rise to a coupling of the form:
\begin{equation}
\label{perturbation}
\hat{U}_{n, S_{z}}(t) = \frac{1}{V} \delta U_{}(\vec{q},\omega) e^{- i \omega t} \sum_{\vec{k}} \, \sum_{\sigma} (\pm 1)^{\sigma}
\, \hat{c}^{\dagger}_{\vec{k} - \vec{q}/2, \sigma} \,  \hat{c}^{}_{\vec{k} + \vec{q}/2, \sigma}
+ h. c.\,,
\end{equation}
where the $+$ sign is for the density, while the minus sign is for the spin density.

The system, at first order in the strength of the perturbation, will respond by generating a density (spin density) modulation $\delta n(\vec{r},t) =
\frac{1}{V} \delta n(\vec{q},\omega) e^{i (\vec{q} \cdot \vec{r} - \omega t)} + c.c $
(and correspondingly $\delta S_z(\vec{r},t)$ for spin density) with:
\begin{equation}
\delta n(\vec{q},\omega) = \chi_{nn}(\vec{q},\omega) \, \delta U(\vec{q},\omega),
\quad  \delta S_z(\vec{q},\omega) = \chi_{S_zS_z}(\vec{q},\omega) \, \delta U(\vec{q},\omega)
\end{equation}
where $\chi_{nn}(\vec{q},\omega)$ and $\chi_{S_zS_z}(\vec{q},\omega)$ are the
density and spin-density response functions of the system.
These functions are related to the density and spin density structure factors
via the celebrated fluctuation-dissipation theorem which, at
zero temperature, reads:
\begin{equation}
\Im\left(  \chi_{nn}(\vec{q},\omega + i 0^{+}) \right) = -\pi n\,  S(\vec{q},\omega), \quad
\Im\left(  \chi_{S_zS_z}(\vec{q},\omega + i 0^{+}) \right) = -\pi n\,  S_s(\vec{q},\omega)\quad .
\end{equation}
For cold gases, the dynamical structure factors can be directly measured with two-photons Bragg spectroscopy.

\COMMENTED{
In this paper we will now present two approaches to compute $S(\vec{q},\omega)$ and
$S_s(\vec{q},\omega)$ starting from the Hamiltonian \eqref{ham_mom}:
the dynamical BCS theory, which yields an approximate expression
for $\chi_{nn}(\vec{q},\omega)$ and $\chi_{S_zS_z}(\vec{q},\omega)$, and the Auxiliary Field Quantum Monte Carlo (AFQMC) method, which provides an unbiased estimation
of the Laplace transforms of $S(\vec{q},\omega)$ and
$S_s(\vec{q},\omega)$ , as we will discuss below.
}

Essentially, dynamical BCS theory attempts to replace the time-dependent Hamiltonian
$\hat{H} + \hat{U}(t)$ with a time-dependent effective Hamiltonian $\hat{H}_{BCS}(t)$
built self-consistently \cite{PhysRev.139.A197,Nozieres,PhysRevA.74.042717}. 
Below we describe the theory briefly, following a short section reviewing
a few aspects of equilibrium BCS theory.

\subsection{Nambu formalism}
We recall that the ground state BCS theory for the homogeneous Fermi gas relies on the following approximation for the interaction term in the Hamiltonian: 
\begin{equation}
\hat{V} \simeq \hat{V}_{BCS} = -  \frac{V |\Delta|^2}{g} +   \sum_{\vec{k}} \Delta^{\star} \, \hat{c}^{}_{-\vec{k}, \downarrow} \hat{c}^{}_{\vec{k}, \uparrow} + \Delta \,  \hat{c}^{\dagger}_{\vec{k}, \uparrow} \hat{c}^{\dagger}_{-\vec{k}, \downarrow}\,,
\end{equation}
where it is assumed that only  singlet zero-momentum Cooper pairs are formed. 
The order parameter $\Delta$ is determined self-consistently via the gap-equation:
\begin{equation}
\Delta = -\frac{g}{V} \sum_{\vec{k}} \langle \hat{c}^{}_{-\vec{k}, \downarrow} \hat{c}^{}_{\vec{k}, \uparrow} \rangle\,,
\end{equation}
where the brackets denote average over the ground state of the one-body Hamiltonian
$\hat{H}_{BCS} = \hat{T} + \hat{V}_{BCS}$.

It is 
useful to introduce the Nambu spinor:
\begin{equation}
\hat{\Psi}^{}(\vec{k}) = 
\left(
\begin{array}{c}
\hat{c}^{}_{\vec{k}, \uparrow} \\
\hat{c}^{\dagger}_{-\vec{k}, \downarrow}
\end{array}
\right), \quad \hat{\Psi}^{\dagger}(\vec{k}) = \left( \hat{c}^{\dagger}_{\vec{k}, \uparrow}, \hat{c}^{}_{-\vec{k}, \downarrow}  \right)\,,
\end{equation}
which allows us to express:
\begin{equation}
\hat{H}_{BCS} = \sum_{\vec{k}} \,  \hat{\Psi}^{\dagger}(\vec{k}) 
\left(
\begin{array}{cc}
\xi(k)  &   \Delta^{\star} \\
\Delta  & -\xi(k)
\end{array}
\right)^{T} \hat{\Psi}^{}(\vec{k}) - \frac{V |\Delta|^2}{g} + \sum_{\vec{k}} \xi(k)\,.
\end{equation}
We will neglect the constant from now on. Moreover, will assume that $\Delta$ is real
and so we will drop the complex conjugation. Note that $\Delta < 0$ in our notation.

The mean-field Hamiltonian can be straightforwardly diagonalized through 
a Bogoliubov transformation into quasi-particles creation and destruction
operators:
\begin{equation}
\hat{\Psi}^{}(\vec{k}) = \mathcal{W}_{\vec{k}} \, \hat{\Phi}^{}(\vec{k}), \quad \hat{\Phi}^{}(\vec{k}) = \left(
\begin{array}{c}
\hat{\alpha}^{}_{\vec{k}} \\
\hat{\beta}^{\dagger}_{-\vec{k}}
\end{array}
\right)
\end{equation}
The transformation matrix can be written in the simple form:
\begin{equation}
\mathcal{W}_{\vec{k}} = \left(
\begin{array}{cc}
u_{k}  &  v_{k} \\
-v_{k}  & u_{k}
\end{array}
\right), \quad u_k = \sqrt{\frac{1}{2}\left( 1 + \frac{\xi(k)}{E(k)}\right)}, \quad v_k = \sqrt{\frac{1}{2}\left( 1 - \frac{\xi(k)}{E(k)}\right)}\,,
\end{equation}
where the quasi-particles dispersion is given by:
\begin{equation}
E(k) = \sqrt{\xi(k)^2 + \Delta^2}\,.
\end{equation}
That is
\begin{equation}
\mathcal{W}^{\dagger}_{\vec{k}} \, \left( e^{(0)}(k) \right)^T  \, \mathcal{W}_{\vec{k}} = \left(
\begin{array}{cc}
E(k)  &   0 \\
0  & -E(k)
\end{array}
\right), \quad 
{\rm with}\ 
  e^{(0)}(k) \equiv \left(
\begin{array}{cc}
\xi(k)  &   \Delta \\
\Delta  & -\xi(k)
\end{array}
\right)\,.
\end{equation}

\COMMENTED{
Since the BCS ground state is annihilated by all the quasi-particle destruction operators,
we easily see that, defining:

\begin{equation}
n^{(0)}_{i,j}(k) = \langle \hat{\Psi}^{\dagger}_i(\vec{k}) ,  \hat{\Psi}^{}_j(\vec{k}) \rangle
\end{equation}
we have:

\begin{equation}
n^{(0)}(k) = \mathcal{W}^{\star}_{\vec{k}} \, \left(
\begin{array}{cc}
0  &   0 \\
0  & 1
\end{array}
\right)
\mathcal{W}^{T}_{\vec{k}}
\end{equation}
}

\subsection{Time-dependent formalism}

Suppose now we switch on a periodic
time-dependent external field with wave-vector
$\vec{q}$ and frequency $\omega$, as defined in \eqref{perturbation},  
coupled to the particle or spin density of the system.
We will focus on the particle density;
the formalism for the spin density case is similar, and simpler.

The central idea is that, at time  $t$, as the system responds
to the perturbation, it develops a time-dependent order parameter self-consistently. The Cooper pairs 
will 
now be allowed to have total momentum $\vec{q}$.
Using Nambu formalism, we can write a time-dependent BCS Hamiltonian
in the form:
\begin{equation}
\hat{H}_{BCS}(t) = \sum_{\vec{k}} \,  \hat{\Psi}^{\dagger}(\vec{k}) 
\left( e^{(0)}(\vec{k}) \right)^T
\hat{\Psi}^{}(\vec{k}) + \sum_{\vec{k}} \,  \hat{\Psi}^{\dagger}(\vec{k} - \vec{q}/2) \,
\left(\delta e (\vec{q},t) \right)^T \, \hat{\Psi}^{}(\vec{k} + \vec{q}/2) + h.c.
\end{equation}
where we have introduced the time-dependent matrix:
\begin{equation}
\delta e (\vec{q},t) = 
\left(
\begin{array}{cc}
\frac{1}{V}\delta U(\vec{q},\omega) e^{-i \omega t}  &   \left( \Delta_{-\vec{q}}(t) \right)^{\star}\\
 \Delta_{\vec{q}}(t)   & -\frac{1}{V} \delta U(\vec{q},\omega) e^{-i \omega t}
\end{array}
\right)
\end{equation}

The matrix $\delta e (\vec{q},t)$ contains both the external perturbation
and the self-consistently generated time-dependent gap function:
\begin{equation}
\Delta_{\vec{q}}(t) = - \frac{g}{V} \sum_{\vec{k}} \langle \hat{c}^{}_{-(\vec{k} - \vec{q}/2),\downarrow} \hat{c}^{}_{\vec{k} + \vec{q}/2),
\uparrow} \rangle
\end{equation}
where the brackets denote an average over the time-dependent ground state of $\hat{H}_{BCS}(t)$.
This dynamical gap equation can be combined with the well known equilibrium gap equation:
\begin{equation}
\frac{1}{g} = \frac{1}{V} \sum_{\vec{k}} \frac{1}{2E(k)}
\end{equation}
to give the conditions:
\begin{equation}
\label{self_consistency}
\begin{split}
& \sum_{\vec{k}} \left( \delta n_{2,1}(\vec{k},\vec{q},t) +  \frac{1}{2E(k)} \Delta_{\vec{q}}(t) \right)  = 0 \\
& \sum_{\vec{k}} \left( \delta n_{1,2}(\vec{k},\vec{q},t) +  \frac{1}{2E(k)} \left( \Delta_{-\vec{q}}(t) \right)^{\star} \right) = 0\,,
\end{split}
\end{equation}
which have to be fulfilled  by  the time dependent fluctuating part of the density matrix:
\begin{equation}
\delta n_{i,j}(\vec{k},\vec{q},t) = \left\langle \hat{\Psi}^{\dagger}_i(\vec{k} - \vec{q}/2) \,  \hat{\Psi}^{}_j(\vec{k} + \vec{q}/2)  \right\rangle\,.
\end{equation}

In order to compute dynamical response functions, the key ingredient is the time derivative:
\begin{equation}
i  \frac{d}{dt}  \hat{\Psi}^{\dagger}_i(\vec{k} - \vec{q}/2) \,  \hat{\Psi}^{}_j(\vec{k} + \vec{q}/2)
= \left[ \hat{\Psi}^{\dagger}_i(\vec{k} - \vec{q}/2) \,  \hat{\Psi}^{}_j(\vec{k} + \vec{q}/2) \, , \, \hat{H}_{BCS}(t) \right]
\end{equation}
which, to first order in the perturbation strength, 
leads to the following kinetic equation for the $2 \times 2$ density matrix 
$\delta n_{\vec{k}}(\vec{q},\omega) = \delta n(\vec{k},\vec{q},t) e^{i \omega t}$:
\begin{equation}
\label{kinetic_equation}
\begin{split}
& \omega \delta n_{\vec{k}}(\vec{q},\omega) =  \delta n_{\vec{k}}(\vec{q},\omega) e^{(0)}(\vec{k} + \vec{q}/2)
- e^{(0)}(\vec{k} - \vec{q}/2) \delta n_{\vec{k}}(\vec{q},\omega) \\
& +  n^{(0)}(\vec{k} - \vec{q}/2) \delta e (\vec{q},\omega) - \delta e (\vec{q},\omega) n^{(0)}(\vec{k} + \vec{q}/2)\,, 
\end{split}
\end{equation}
where 
$n^{(0)}_{i,j}(\vec{k}) = \left\langle \hat{\Psi}^{\dagger}_i(\vec{k}) \,  \hat{\Psi}^{}_j(\vec{k})  \right\rangle$ 
is the equilibrium density matrix.

Equations~\eqref{kinetic_equation}, complemented with the conditions \eqref{self_consistency}, can be solved to 
obtain the density-density response function.
We skip the simple but lengthy algebra here and give directly the final result:
\begin{equation}
\chi_{nn}
 = -\frac{1}{V} \left\{  I^{\prime\prime} + \frac{\Delta^2}{I_{11} I_{22}
- \omega^2 I_{12}}
\left(2\omega^2  
I \, I_{12} \, I^{\prime}  -\omega^2 I_{22}  I^{\prime \, 2}  - I^2 \, I_{11}   \right)  \right\}\,,
\end{equation}
where
\COMMENTED{
\begin{equation}
I''(\vec{q},\omega) = \sum_{\vec{k}} \frac{E_{+} + E_{-}}{\left( E_{+} + E_{-} \right)^2 - \omega^2} \left( \frac{E_{-}E_{+}-\xi_{-}\xi_{+}+\Delta^2}{E_{-}E_{+}}\right)   
\end{equation}
\begin{equation}
I(\vec{q},\omega) = \sum_{\vec{k}} \frac{E_{+} + E_{-}}{\left( E_{+} + E_{-} \right)^2 - \omega^2} \left( \frac{\xi_{+} + \xi_{-}}{E_{-}E_{+}}\right)   
\end{equation}
\begin{equation}
I'(\vec{q},\omega) = \sum_{\vec{k}} \frac{E_{+} + E_{-}}{\left( E_{+} + E_{-} \right)^2 - \omega^2} \left( \frac{1}{E_{-}E_{+}}\right)   
\end{equation}
\begin{equation}
I_{12}(\vec{q},\omega) = \sum_{\vec{k}}\frac{1}{\left( E_{+} + E_{-} \right)^2 - \omega^2}  \left(  \frac{E_{+} \xi_{-} + E_{-} \xi_{+}}{E_{-} E_{+} }\right)  
\end{equation}
\begin{equation}
I_{11}(\vec{q},\omega) = \sum_{\vec{k}}\frac{E_{+} + E_{-}}{\left( E_{+} + E_{-} \right)^2 - \omega^2} \left(\frac{E_{-} E_{+} + \xi_{-}\xi_{+} + \Delta^2}{E_{-} E_{+}} \right)   - \frac{1}{E}  
\end{equation}
\begin{equation}
I_{22}(\vec{q},\omega) = \sum_{\vec{k}}\frac{E_{+} + E_{-}}{\left( E_{+} + E_{-} \right)^2 - \omega^2} \left(\frac{E_{-} E_{+} + \xi_{-}\xi_{+} - \Delta^2}{E_{-} E_{+}} \right)   - \frac{1}{E}  
\end{equation}
}
\begin{equation}
\begin{split}
&I''(\vec{q},\omega) = \sum_{\vec{k}} \frac{E_{+} + E_{-}}{\left( E_{+} + E_{-} \right)^2 - \omega^2} \left( \frac{E_{-}E_{+}-\xi_{-}\xi_{+}+\Delta^2}{E_{-}E_{+}}\right)   \\
&I(\vec{q},\omega) = \sum_{\vec{k}} \frac{E_{+} + E_{-}}{\left( E_{+} + E_{-} \right)^2 - \omega^2} \left( \frac{\xi_{+} + \xi_{-}}{E_{-}E_{+}}\right)   \\
& I'(\vec{q},\omega) = \sum_{\vec{k}} \frac{E_{+} + E_{-}}{\left( E_{+} + E_{-} \right)^2 - \omega^2} \left( \frac{1}{E_{-}E_{+}}\right)   \\
& I_{12}(\vec{q},\omega) = \sum_{\vec{k}}\frac{1}{\left( E_{+} + E_{-} \right)^2 - \omega^2}  \left(  \frac{E_{+} \xi_{-} + E_{-} \xi_{+}}{E_{-} E_{+} }\right)  \\
& I_{11}(\vec{q},\omega) = \sum_{\vec{k}}\frac{E_{+} + E_{-}}{\left( E_{+} + E_{-} \right)^2 - \omega^2} \left(\frac{E_{-} E_{+} + \xi_{-}\xi_{+} + \Delta^2}{E_{-} E_{+}} \right)   - \frac{1}{E}  \\
& I_{22}(\vec{q},\omega) = \sum_{\vec{k}}\frac{E_{+} + E_{-}}{\left( E_{+} + E_{-} \right)^2 - \omega^2} \left(\frac{E_{-} E_{+} + \xi_{-}\xi_{+} - \Delta^2}{E_{-} E_{+}} \right)   - \frac{1}{E}  
\end{split}
\end{equation}
We use the notation $E_{+} = E(\vec{k} + \vec{q}/2)$ and $E_{-} = E(\vec{k} - \vec{q}/2)$
for $E$ and $\xi$ for simplicity.
The time-dependent
self-consistent order parameter in the case of a perturbation coupled to the
particle density is given by:
\begin{equation}
\delta \Delta_{\vec{q}}(t)  = \frac{e^{-i\omega t} \, \delta U(\vec{q},{\omega}) \, \Delta}{V \, \left( I_{11} I_{22} - \omega^2 I^2_{12}\right)}\left( -\omega I \, I_{12} + I \, I_{11}  -\omega^2 I' \, I_{12} + \omega I' \, I_{22} \right)
\end{equation}

The computation of the spin density response function leads to a much simpler result:
\begin{equation}
\chi_{S_zS_z}
 = -\frac{1}{V} I_s^{\prime\prime} \,,
\end{equation}
where
\begin{equation}
I''_s(\vec{q},\omega) = \sum_{\vec{k}} \frac{E_{+} + E_{-}}{\left( E_{+} + E_{-} \right)^2 - \omega^2} \left( \frac{E_{-}E_{+}-\xi_{-}\xi_{+} -\Delta^2}{E_{-}E_{+}}\right)\,.   
\end{equation}

\begin{figure}[ptb]
\begin{center}
\includegraphics[width=10cm, angle = 270]{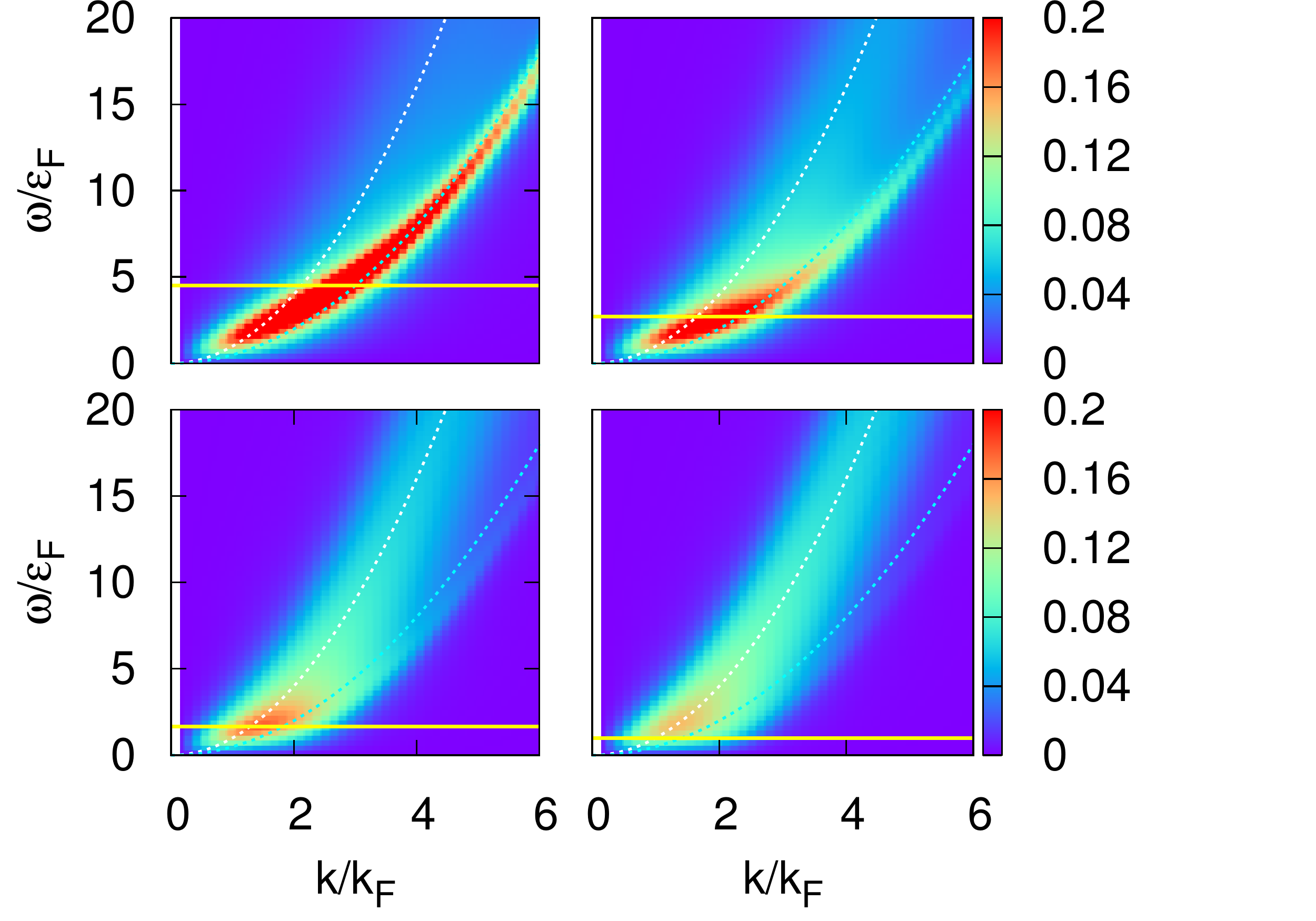}
\caption{(color online) Color plot of the dynamical structure factor $S(\vec{q},\omega)$ (units $1/\varepsilon_F = (2m/\hbar^2) (1/ 2 \pi n )$) of
a $2D$ Fermi gas for four values of the interaction parameter:  $\log(k_F a) = 0.0$ (top left panel), $0.5$ (top right panel),
$1.0$ (bottom left panel) and $1.5$ (bottom right panel).
We also show an horizontal line at $2|\Delta|$, the threshold to break Cooper pairs,
the free atoms dispersion $e_a(q) = \hbar^2|\vec{q}|^2/ 2 m$, and the free molecule dispersion $e_m(q) = \hbar^2|\vec{q}|^2/ 4 m$.
 }
\label{fig:dBCS_dens}
\end{center}
\end{figure}

\begin{figure}[ptb]
\begin{center}
\includegraphics[width=10cm, angle = 270]{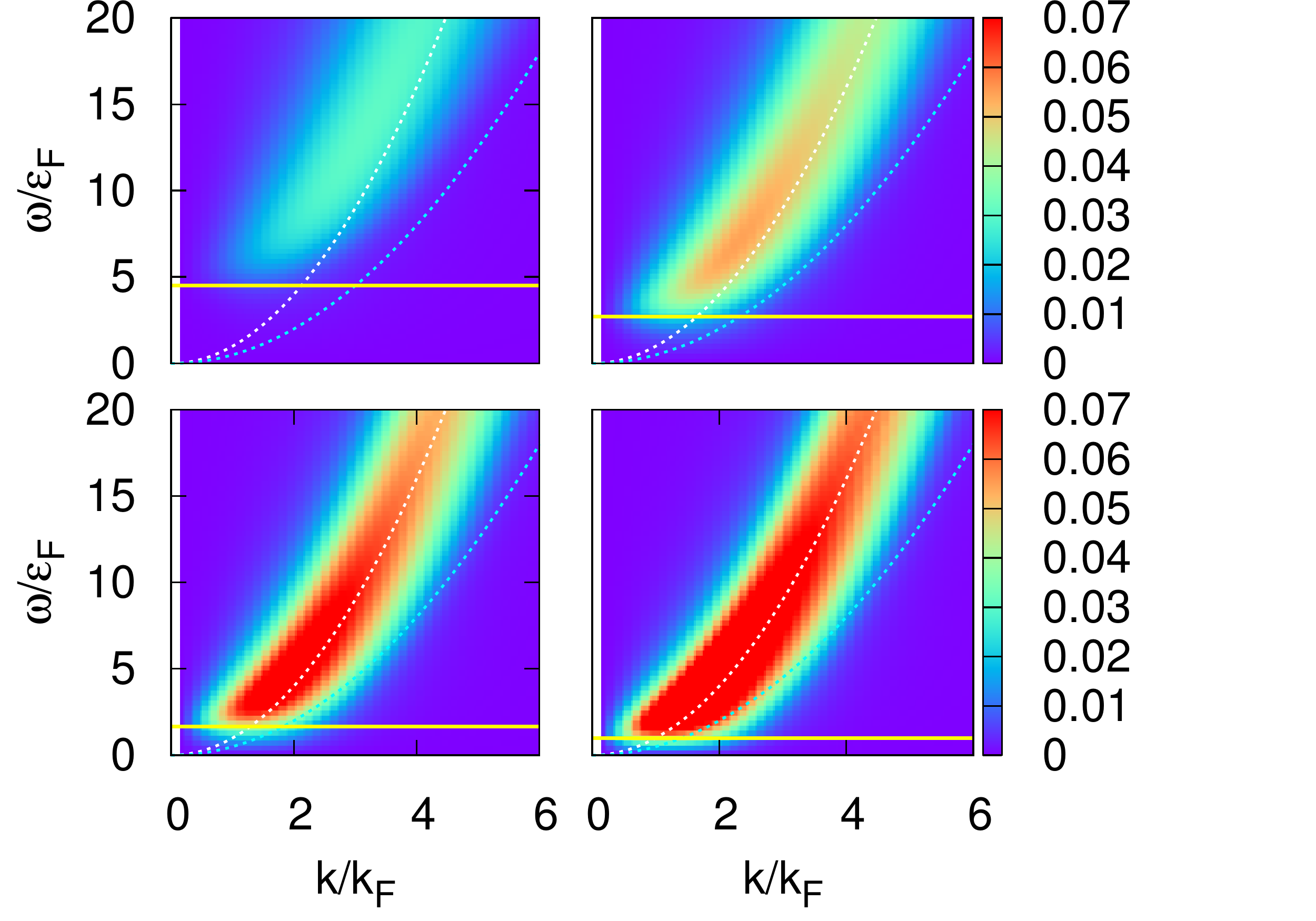}
\caption{(color online) Color plot of the spin dynamical structure factor $S_s(\vec{q},\omega)$ 
(units $1/\varepsilon_F$) of
a $2D$ Fermi gas. The setup is the same as in Fig.~\ref{fig:dBCS_dens}.
}
\label{fig:dBCS_spin}
\end{center}
\end{figure}

The results for the dynamical structure factor for the density $S(\vec{q},\omega)$ and the spin density $S_s(\vec{q},\omega)$ are shown in Figs.~\ref{fig:dBCS_dens} and \ref{fig:dBCS_spin} respectively for four values of the
interaction parameter,  
from the BEC regime $\log(k_F a) = 0.0$ through  
to the BCS regime of $\log(k_F a) =1.5$. 

At $\log(k_F a) = 0.0$ 
the system is expected
to behave as a gas of molecules. This is confirmed by the high momentum behavior
$\hbar^2|\vec{q}|^2/ 2 m >> |\Delta|$ 
of $S(\vec{q},\omega)$, where we have a spectrum of free molecules $S(\vec{q},\omega)
\simeq \delta ( \omega - e_m(\vec{q}))$, with $e_m(\vec{q}) =  \hbar^2|\vec{q}|^2/ 4 m$ containing the mass of the molecules ($2m$). 
We observe that this behavior
is not present is the spin structure factor $S_s(\vec{q},\omega)$: in order to
create a modulation in spin density, it is necessary to first break a Cooper pair.
This is the reason 
that the appreciable values for $S_s(\vec{q},\omega)$ lie 
above
the line $2 |\Delta|$, which is the 
estimated energy cost
of breaking a molecule. 

On the other side, in the BCS regime at $\log(k_F a) = 1.5$ , the dynamical
structure factors are more similar to the non-interacting results. The spectrum
is centered around the free atom dispersion $e_{a}(\vec{q}) = \hbar^2|\vec{q}|^2/ 2 m$,
with a broadening due to the particle-hole continuum.

We also see that the 
theory predicts
a smooth interpolation between the two regimes, 
 consistent with a BEC-BCS crossover. 
 From comparison with AFQMC calculations, which we discuss next, it is seen
 that the results  are in reasonably good agreement, although 
 quantitative differences exist.

\COMMENTED
{
Note that, by construction:

\begin{equation}
\delta n_{1,2}(\vec{k},\vec{q},t) = \left(\delta n_{2,1}(\vec{k},-\vec{q},t)\right)^{\star}
\end{equation}
that is:

\begin{equation}
\left(\delta n_{\vec{k}}(\vec{q},\omega)\right)_{1,2} = \left( \left( \delta n_{\vec{k}}(-\vec{q},-\omega)\right)_{2,1} \right)^{\star}
\end{equation}

Let's start from an Hamiltonian:
\begin{equation}
\hat{H}(t) = \hat{H} + \hat{U}(t)
\end{equation}
where $\hat{H}$ is the Hamiltonian for the cold gas \eqref{ham}, while
$\hat{U}(t)$ is a perturbation, with the simple interpretation of a periodic
time dependent external field:

\begin{equation}
\hat{U}(t) = \delta U \left( e^{i \omega t} \hat{\rho}_{\vec{q}} + e^{-i \omega t} \hat{\rho}_{-\vec{q}} \right)
\end{equation}
coupled to the particle or spin density $\hat{\rho}_{\vec{q}}$ of the system.

We find very useful to introduce Nambu spinor:

\begin{equation}
\hat{\Psi}^{}(\vec{k}) = 
\left(
\begin{array}{c}
\hat{c}^{}_{\vec{k}, \uparrow} \\
\hat{c}^{\dagger}_{-\vec{k}, \downarrow}
\end{array}
\right), \quad \hat{\Psi}^{\dagger}(\vec{k}) = \left( \hat{c}^{\dagger}_{\vec{k}, \uparrow}, \hat{c}^{}_{-\vec{k}, \downarrow}  \right)
\end{equation} 
}

\section{Quantum Monte Carlo Approach}
\label{AFQMC}

Because of the mean-field nature of the approximations involved in dynamical BCS theory,
it is not clear \emph{a priori} how reliable the results are, 
especially  in strongly interacting regimes.
Quantum Monte Carlo (QMC) approaches provide an
alternative. In particular, 
for the spin-balanced cold Fermi gas there is no fermion sign problem
in the auxiliary-field Quantum Monte Carlo (AFQMC) method \cite{AFQMC-lecture-notes-2013,hubbard_benchmark,PhysRevB.94.085103,PhysRevB.88.125132}.
%
The AFQMC method is able to provide unbiased, numerically exact results 
for any observable on the ground state of the cold Fermi gas \cite{Hao-2DFG,PhysRevA.84.061602}.
We have recently showed \cite{ettoreGAP,Ettore-2DFG} that it is possible to reach beyond static properties and compute 
imaginary-time correlation functions such as
\begin{equation}
\label{dynamical_general}
F(\vec{q},\tau) = \frac{1}{N} \frac{ \langle \Psi_0  | \, \hat{n}_{\vec{q}}  \, e^{-\tau (\hat{H} - E_0) } \, \hat{n}_{-\vec{q}} \, | \Psi_0 \rangle }
{\langle \Psi_0 \, |  \, \Psi_0 \rangle }
\end{equation}
where $N$ is the number of particles and $\hat{n}_{\vec{q}}$ the Fourier
component of the density (or spin density) fluctuation operator.
The exact zero-temperature relation:
\begin{equation}
\label{dynamical_density_tau}
F(\vec{q},\tau) = \int_{0}^{+\infty} d\omega e^{-\tau \omega} S(\vec{q},\omega)
\end{equation}
allows us to then 
obtain predictions about the dynamical structure factor 
(density and spin density) of the system using analytic continuation techniques.
%
In the following we will introduce the AFQMC method and discuss how
to efficiently compute $F(\vec{q},\tau)$ for the Fermi gas system.

\subsection{AFQMC formalism and static properties}
We will introduce the basic notations of the
methodology using the attractive Hubbard
model Hamiltonian \eqref{ham_latt} 
in Sec.~\ref{hcgr}.
The AFQMC methodology relies on the following \cite{AFQMC-lecture-notes-2013}: 
 \begin{equation}
| \, \Psi_0 \rangle \propto \lim_{\beta \to +\infty} e^{-\beta ( \hat{H} - E_0)} |\phi_T\rangle\,,
\end{equation}
where $E_0$ is an estimate of the ground state energy,
and  $|\phi_T\rangle$ a trial wave function 
which is not orthogonal to the 
many-body ground state $ | \,\Psi_0 \, \rangle$. 

Trotter-Suzuki breakup together with Hubbard-Stratonovich transformation provides
the following: 
 \begin{equation}
 \label{propagator}
e^{-\beta ( \hat{H} - E_0)}  = \left(e^{-\delta\tau ( \hat{H} - E_0)} \right)^{M} \simeq  \left(\int d{\bf{x}} p({\bf{x}}) \hat{B}({\bf{x}})\right)^{M}\,, 
\end{equation}
which becomes exact in the limit $M\to \infty$ ($\delta\tau = \beta/M$
is a {\it{time-step}}). At each time slice, 
there is one set of auxiliary-fields,
${\bf{x}} = (x_1, \dots, x_{\mathcal{N}_s})$, 
which are a discrete set of Ising fields on the lattice. 
$\hat{B}({\bf{x}})$ is a one-particle propagator, and 
the function $p({\bf{x}})$ is a discrete 
probability density.

A key point of the methodology is that $\hat{B}(\bf{x})$ 
in  Eq.~\eqref{propagator},
which is the result of the HS transformation, is the exponential of a one-body propagator; its application on a Slater determinant $|\phi\rangle$ results in  another Slater determinant $|\phi'\rangle$, given in matrix form by
\begin{equation}
\mathcal{B}(\bf{x}) \Phi =\Phi'\,,
 \label{eq:Thouless}
\end{equation}
where 
$\Phi=\Phi_\uparrow\otimes \Phi_\downarrow$, with $\Phi_\sigma$ being the $\mathcal{N}_s\times N_\sigma$ matrix 
containing the spin-$\sigma$ orbitals of the Slater determinant $|\phi\rangle$,
and similarly for $|\phi'\rangle$.

The standard path-integral AFQMC method allows us to evaluate ground state
expectation values:
\begin{equation}
\label{exp_val}
\langle \hat{O} \rangle = \frac{ \langle \Psi_0  | \, \hat{O} \, | \Psi_0 \rangle }
{\langle \Psi_0 \, | \, \Psi_0 \rangle }
\end{equation}
by casting them in the integral form:
\begin{equation}
\label{exp_val2}
\langle \hat{O} \rangle = \int d{\bf{X}} \, \mathcal{W}({\bf{X}}) \, \mathcal{O}({\bf{X}})\,.
\end{equation}
In Eq.~\eqref{exp_val2}, ${\bf X} = ({\bf x}(1), \dots, {\bf x}(M))$ denotes a {\it{path}} in auxiliary-field 
space. 
Choosing the trial wave function  $|\phi_T\rangle$ as a single Slater determinant and 
introducing the notation:
\begin{equation}
\label{phi_L}
\langle \phi_L  | \,  = \langle \phi_T  | \,\hat{B}({\bf x}(M)) \dots \hat{B}({\bf x}(l)) 
\end{equation}
and
\begin{equation}
\label{phi_R}
| \phi_R \rangle = 
\hat{B}({\bf x}(l-1)) \dots \hat{B}({\bf x}(1))  \,| \phi_T \rangle\,,
\end{equation}
we have:
\begin{equation}
 \mathcal{W}({\bf{X}}) \propto \, \langle \phi_L \, |   \, \phi_R \rangle \,  \prod_{i=1}^{M} p({\bf{x}}(i))\,,
\end{equation}
while the estimator is
\begin{equation}
\label{static_estimator}
\mathcal{O}({\bf{X}}) = \frac{ \langle \phi_L  | \, \hat{O} \, | \phi_R \rangle }
{\langle \phi_L \, | \, \phi_R \rangle }\,.
\end{equation}

For attractive interaction and $N_\uparrow=N_\downarrow$, 
$ \mathcal{W}({\bf{X}}) $ remains non-negative for all possible  auxiliary-field path configurations.
The calculation
is sign-problem free, allowing us to obtain exact results.
%
We use a Metropolis sampling of the paths, exploiting a force 
bias \cite{Hao-2DFG,AFQMC-lecture-notes-2013} that allows high acceptance ratio in the updates. 
The infinite variance problem is eliminated
with a bridge link approach \cite{Hao-inf-var}.

\subsection{Dynamical properties}
\label{ssec:method-dynamical}

The AFQMC methodology allows us to also compute 
dynamical correlation functions in imaginary-time
at zero temperature:
\begin{equation}
\label{dynamical}
f(\tau) = \frac{ \langle \Psi_0  | \, \hat{O} \, e^{-\tau (\hat{H} - E_0) } \, \hat{O}^{\dagger} \, | \Psi_0 \rangle }
{\langle \Psi_0 \, |  \, \Psi_0 \rangle }\,,
\end{equation}
where $\hat{O}$  can be any operator. We will focus here
on one-body
operators such as the particle density or the spin density.

The imaginary-time propagator between the operators  $\hat{O}$ and $\hat{O}^{\dagger}$ 
can be expressed using Eq.~\eqref{propagator}. 
We insert an extra segment to the path:
a number $N_{\tau} = \tau/\delta\tau$ of
{\it{time-slices}}, ${\bf \tilde{x}}(1), \dots , {\bf \tilde{x}}(N_{\tau})$.
The static estimator in Eq.~\eqref{static_estimator} is replaced 
by the following dynamical estimator:
\begin{equation}
\label{dynamical estimator}
f({\bf{X}},\tau) = \frac{ \langle \phi_L  | \,  \hat{O} \, \hat{B}({\bf{\tilde{x}}}(N_{\tau})) \dots  \hat{B}({\bf{\tilde{x}}}(1))  \,\,\hat{O}^{\dagger} | \phi_R \rangle }
{\langle \phi_L \, |  \hat{B}({\bf{\tilde{x}}}(N_{\tau})) \dots  \hat{B}({\bf{\tilde{x}}}(1)) \,| \, \phi_R \rangle }\,.
\end{equation}
Below we will write $\hat{B}_i$ instead of  $\hat{B}({\bf \tilde{x}}(i))$ for notational simplicity.

Standard approaches \cite{assaad_prb,hirsch-stable,jel_dyn_method,jel_dyn_method2} to evaluate the expression in Eq.~(\ref{dynamical estimator}) require
computational cost of $\mathcal{O}(\mathcal{N}^3_s)$. 
We have recently introduced an approach  \cite{ettoreGAP} which allows computation of the 
matrix elements of 
dynamical Green functions and (spin) density correlation functions 
with computational cost of $\mathcal{O}(\mathcal{N}_s\,N^2)$.
For dilute systems such as the ultracold Fermi 
gas that we are concerned with  here,  the number of particles is 
significantly smaller than the number of lattice sites, $N \ll \mathcal{N}_s $. The approach thus offers a significant advantage which enables us to reach the realistic limits.
Since the estimator \eqref{dynamical estimator} is computed
sampling the same probability distribution as for 
static properties \eqref{static_estimator}, a finite imaginary time $\tau > 0$ 
does not introduce any bias.
We will now sketch the method, which explicitly propagates
fluctuations during the random walk. 

\COMMENTED{
we take advantage
of a computation scheme to compute \eqref{dynamical estimator}
introduced by ourselves \cite{ettoreGAP}
for dynamical Green functions and (spin) density correlation functions,
whose computational cost, for matrix elements, is $\mathcal{O}(\mathcal{N}_s\,\mathcal{N}_p^2)$ 
versus the typical $\mathcal{O}(\mathcal{N}^3_s)$ of well established approaches
}

\subsection{Two-body correlation functions}
Let us focus on the estimator:
\begin{equation}
\label{dynamical estimator_new_gen_tb}
n({\bf{X}},\tau) =\frac{ \langle \phi_L  | \, \hat{n}_{j,\sigma'}  \,\hat{B}_{N_{\tau}} \dots \hat{B}_1  \,\,\hat{n}_{i,\sigma}  | \phi_R \rangle }
{\langle \phi_L \, | \,\hat{B}_{N_{\tau}} \dots \hat{B}_1  \,\,| \, \phi_R \rangle }
\end{equation}
where, as usual, $\hat{n}_{i,\sigma} = \hat{c}^{\dagger}_{i,\sigma}  \hat{c}^{}_{i,\sigma} $
is the fermion spin-resolved density operator.

It is straightforward to show
\begin{equation}
\label{smart}
 \hat{n}_{i,\sigma} = \frac{e^{\hat{n}_{i,\sigma}} - 1}{e-1}\,.
\end{equation}
This implies that, if $| \phi_R \rangle = \hat{c}^{\dagger}_{|u_1,\uparrow\rangle} \dots \hat{c}^{\dagger}_{|u_{N_\uparrow},\uparrow\rangle}
 \hat{c}^{\dagger}_{|v_1,\downarrow\rangle} \dots \hat{c}^{\dagger}_{|v_{N_\downarrow},\downarrow\rangle} |0\rangle$,
 the numerator in Eq.~\eqref{dynamical estimator_new_gen_tb} can be viewed as the propagation of two Slater determinants:
\begin{equation}
 \hat{n}_{i,\uparrow}| \phi_R \rangle = \frac{ | \phi'_R(i) \rangle -  | \phi_R \rangle}{e-1}
\end{equation}
where:
\begin{equation}
| \phi_R'(i) \rangle =
\hat{c}^{\dagger}_{|e^{\hat{n}_{i,\uparrow}}u_1,\uparrow\rangle} \dots \hat{c}^{\dagger}_{|e^{\hat{n}_{i,\uparrow}}u_{N_\uparrow},\uparrow\rangle}
 \hat{c}^{\dagger}_{|v_1,\downarrow\rangle} \dots \hat{c}^{\dagger}_{|v_{N_\downarrow},\downarrow\rangle} |0\rangle\,.
\end{equation}
Consequently, Eq.~\eqref{dynamical estimator_new_gen_tb} can be broken into
two pieces:
\begin{equation}
\label{dynamical estimator_new_gen_tb-broken}
n({\bf{X}},\tau) = \frac{1}{e-1} \left(n_1({\bf{X}},\tau) - n_2({\bf{X}},\tau)\right)\,,
\end{equation}
which can be  expressed as:
\begin{equation}
n_1({\bf{X}},\tau) = \frac{ \langle \phi_L  | \, \hat{n}_{j,\sigma'} \hat{B}  \,\,  | \phi'_R(i) \rangle }
{\langle \phi_L \, | \hat{B}  \,\,| \, \phi'_R (i)\rangle }
\,\, \frac{ \langle \phi_L  | \, \hat{B}  \,\,  | \phi'_R(i) \rangle }
{\langle \phi_L \, | \hat{B}  \,\,| \, \phi_R \rangle }
\end{equation}
and
\begin{equation}
n_2({\bf{X}},\tau)  =  \frac{ \langle \phi_L  | \, \hat{n}_{j,\sigma'} \hat{B}  \,\,  | \phi_R \rangle }
{\langle \phi_L \, | \hat{B}  \,\,| \, \phi_R \rangle }\,.
\end{equation}
Both $n_1({\bf{X}},\tau)$ and $n_2({\bf{X}},\tau)$ can be calculated 
with straightforward manipulations of Slater determinants.
The average of $n({\bf{X}},\tau)$ over an ensemble of paths
in the configuration space of the auxiliary field provides the estimation
of the density or spin-density imaginary time correlation function.

\begin{figure}[ptb]
\begin{center}
\includegraphics[width=8cm, angle = 270]{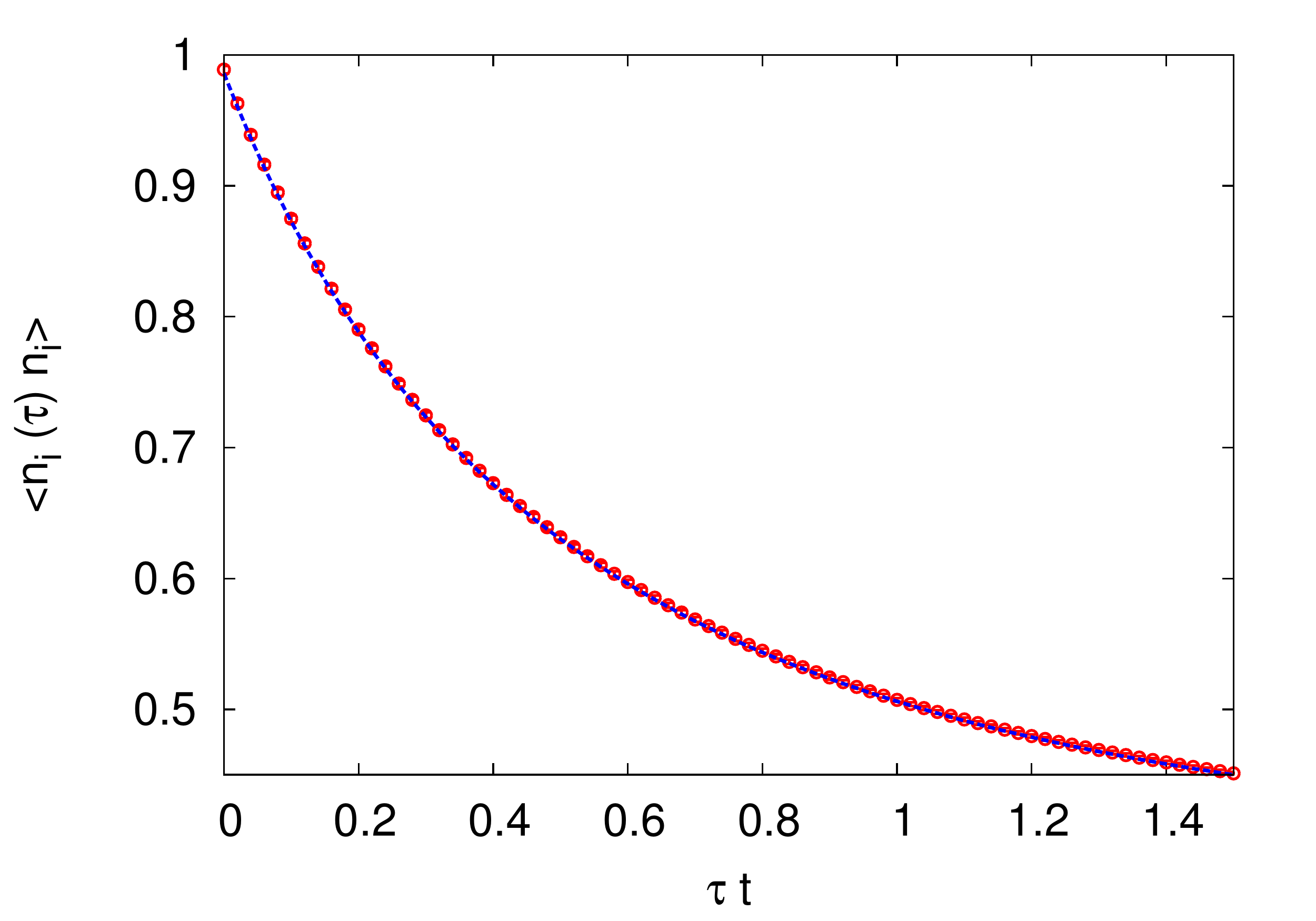}
\caption{(color online) Imaginary-time on-site density-density correlation
function $\langle \hat{n}_{i} \, e^{-\tau(\hat{H} - E_0)} \, \hat{n}_{i}  \rangle$ for the lattice Hamiltonian in Eq.~\eqref{ham_latt}
on a $4 \times 4$ lattice hosting $10$ particles
with $g_{\mathcal{L}}/t = 4$. Comparison between the QMC computation (circles)
and the exact diagonalization result (dotted line).
AFQMC errors bars are shown but are smaller than symbol size.}
\label{fig:dBCS_vs_ED}
\end{center}
\end{figure}

In Fig.~\ref{fig:dBCS_vs_ED} we show a comparison between AFQMC and
exact diagonalization results. We compute the same site density-density correlation
in imaginary time for a $4 \times 4$ lattice hosting $N=10$ particles
with $g_{\mathcal{L}}/t = 4$. 
Excellent  agreement is seen, 
providing a strong calibration of the AFQMC algorithm.

\section{Comparison between dynamical BCS theory and AFQMC}
\label{COMP}

\begin{figure}[ptb]
\begin{center}
\includegraphics[width=10cm, angle = 270]{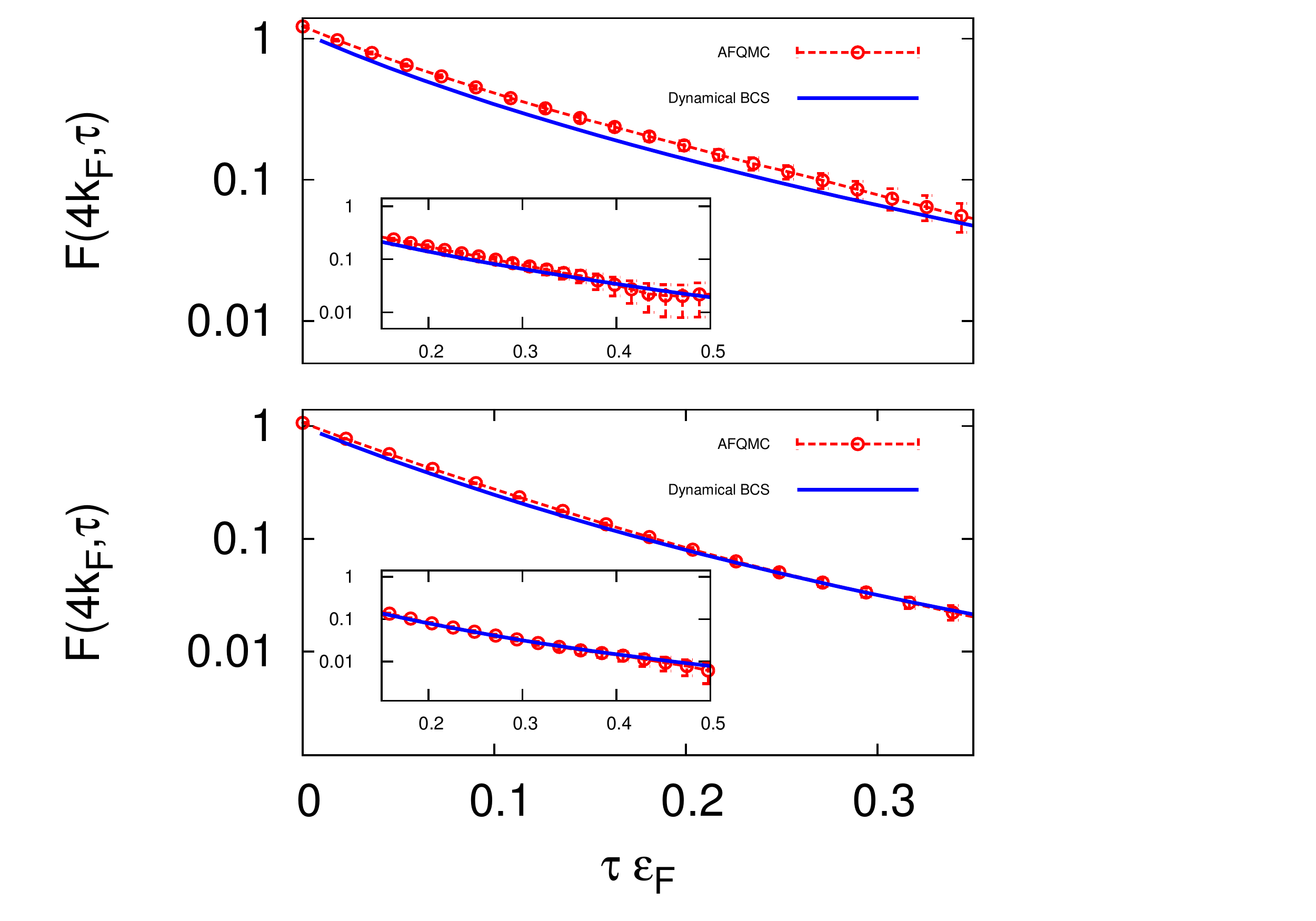}
\caption{(color online) Comparison of imaginary-time density-density correlation,  $F(\vec{q},\omega)$
at $q = 4 k_F$,
obtained from dynamical
BCS theory (blue full line) and AFQMC (red open circles).
The top panel is in the crossover regime,
at $\log(k_F a) = 0.5$, while the bottom panel is in the BCS regime,
at $\log(k_F a) = 1.5$. The insets show the comparison for larger
values of the imaginary time. The imaginary time window considered is large
enough to capture the behavior of the density fluctuation mode at $q = 4 k_F$.}
\label{fig:dBCS_vs_QMC}
\end{center}
\end{figure}

With the algorithm described above, we can compute unbiased imaginary-time correlation functions $F(\vec{q},t)$ for the 2D Fermi gas. The connection with
dynamical structure factors is provided by the relation in Eq.~\eqref{dynamical_density_tau},
which has to be inverted to estimate $S(\vec{q},\omega)$.
This is an analytic continuation problem, which can be delicate.
However, a major advantage here is that the quality of the imaginary-time data is very high.
State of art analytic continuation approaches  have been
shown to provide very accurate estimates under such circumstances, for instance in the realm of cold
bosonic systems \cite{giftREV}.
A comprehensive study 
of the dynamical structure factors from AFQMC for the 2D gas will be 
presented elsewhere \cite{Ettore-2DFG}.

Here we 
directly compare the result from 
dynamical
BCS theory transformed into imaginary time domain with the exact AFQMC
results. We stress that
the direct Laplace transform, from frequency domain to imaginary time 
domain, can be performed with arbitrary accuracy, while the inverse
transformation is much more challenging. 
In order to assess the quality of the dynamical BCS theory against
unbiased AFQMC results, we choose a system of $N = 18$ particles
moving on a lattice with $25 \times 25$ sites, and compute $F(\vec{q},t)$.
We choose a high-momentum transfer of $q = 4 k_F$, which provides an interesting
probe of the BCS-BEC crossover when 
the atomic spectrum 
evolves into the  molecular spectrum. 

In Fig.~\ref{fig:dBCS_vs_QMC} we present the comparisons in the BCS regime
and in the crossover regime. We see that, in the BCS regime at $\log(k_F a) = 1.5$,  dynamical BCS theory prediction
is compatible with the exact result, except for a narrow window at small imaginary time,
where all the excited states of the microscopic Hamiltonian play an important role.
On the other hand, in the crossover regime at $\log(k_F a) = 0.5$, the discrepancy
is larger, although agreement is still restored in the large imaginary time region.
This suggests that many-body correlations
beyond dynamical mean field have important effects on response functions
in the strongly-correlated BEC-BCS crossover.
Considering sum rules, we observe that the $0$th moment, the static structure factor (that is the value of the correlation function at $\tau t = 0$)  is significantly different between the two approaches.
On the other hand, the dynamical BCS theory strictly imposes
the $1$st moment sum rule ("f sum rule''), which is also
satisfied by the QMC result within statistical uncertainties.

Finally, we comment that the dynamical BCS theory calculations here provide a rigorous route to benchmark analytic continuation methods. By applying an analytic continuation method to the dynamical BCS data in the imaginary-time domain, one could directly compare the results against the corresponding results in the frequency domain. Together with the availability of exact AFQMC results in the imaginary-time domain here, the problem of the Fermi gas provides an excellent potential benchmark problem for methods to treat dynamical properties in strongly interacting many-fermion systems.

\section{Conclusions}
\label{CONC}
We have studied the response functions, and more precisely
dynamical structure factors, of a two-dimensional cold gas with
zero-range attractive interactions at zero-temperature. We computed $S(\vec{q},\omega)$ within the framework of dynamical BCS theory, which provides explicit approximate expressions involving integrals over momentum space. We also described an efficient algorithm to compute imaginary-time density and spin correlations 
with AFQMC for
the dilute gases, and used the results to benchmark dynamical BCS theory.

The results of dynamical BCS theory show a 
spectrum of density fluctuations made of a fermionic particle-hole continuum superimposed
to a gapless bosonic mode which, at high momentum, gives the response of a gas of free molecules on the BEC side of the crossover. The weight of such spectrum naturally decreases as we move towards the BCS regime, where the systems is similar to an ordinary superconductor.

The spin density fluctuation spectrum has a simpler structure. It displays a gap corresponding to the energy needed to break a Cooper pair, which is necessary to induce a spin density modulation at
zero-temperature. At higher energies, the spectrum displays the typical
particle-hole continuum.

We have discussed the importance and outlook of AFQMC calculations for imaginary-time correlations
and response functions.
Our AFQMC  results  are used to assess the accuracy of dynamical BCS theory.
Discrepancies arise in the
BEC-BCS crossover, although many qualitative features appear 
to be correctly described by dynamical BCS theory. Outside the strongly correlated regime, the theory is
accurate and captures both the fermionic particle-hole
excitations and the bosonic dynamics.
The smaller differences in this regime tend to be 
more pronounced at  shorter imaginary-times. 
A very interesting perspective will be the study
of the possibility to use effective parameters\cite{PhysRevB.94.235119} in the dynamical
BCS theory to improve the agreement with exact results.

\begin{acknowledgements}
We gratefully acknowledge support from the National Science Foundation (NSF) 
under grant number DMR-1409510 and from 
 the Simons Foundation. 
 The calculations were carried out at the Extreme Science and Engineering Discovery Environment (XSEDE), which is supported by NSF grant number ACI-1053575
 and the computational facilities at the College of William and Mary. 

\end{acknowledgements}



\end{document}